# 3D PET/CT Tumor Lesion Segmentation Based on nnUNet with GCN Refinement

**Hengzhi Xue[1], Qingqing Fang[1], Yudong Yao[1,2] and Yueyang Teng[1,3]**

[1] College of Medicine and Biological Information Engineering, Northeastern University, Shenyang 110004, China  
[2] Department of Electrical and Computer Engineering, Steven Institute of Technology, Hoboken, NJ 07102, USA  
[3] Key Laboratory of Intelligent Computing in Medical Image, Ministry of Education, Shenyang, 110169, China

E-mail: tengyy@bime.neu.edu.cn

**Abstract**

Whole-body PET/CT scan is an important tool for diagnosing various malignancies (e.g., malignant melanoma, lymphoma, or lung cancer), and accurate segmentation of tumors is a key part for subsequent treatment. In recent years, CNN-based segmentation methods have been extensively investigated. However, these methods often give inaccurate segmentation results, such as over-segmentation and under-segmentation. Therefore, to address such issues, we propose a post-processing method based on a graph convolutional neural network (GCN) to refine inaccurate segmentation parts and improve the overall segmentation accuracy. Firstly, nnUNet is used as an initial segmentation framework, and the uncertainty in the segmentation results is analyzed. Certainty and uncertainty nodes establish the nodes of a graph neural network. Each node and its 6 neighbors form an edge, and 32 nodes are randomly selected for uncertain nodes to form edges. The highly uncertain nodes are taken as the subsequent refinement targets. Secondly, the nnUNet result of the certainty nodes is used as label to form a semi-supervised graph network problem, and the uncertainty part is optimized through training the GCN network to improve the segmentation performance. This describes our proposed nnUNet-GCN segmentation framework. We perform tumor segmentation experiments on the PET/CT dataset in the MICCIA2022 autoPET challenge. Among them, 30 cases are randomly selected for testing, and the experimental results show that the false positive rate is effectively reduced with nnUNet-GCN refinement. In quantitative analysis, there is an improvement of 2.12 % on the average Dice score, 6.34 on 95 % Hausdorff Distance (HD95), and 1.72 on average symmetric surface distance (ASSD). The quantitative and qualitative evaluation results show that GCN post-processing methods can effectively improve tumor segmentation performance.

## 1. Introduction

Multi-modality positron emission tomography/computed tomography (PET/CT) scan is an indispensable part for early tumor screening, diagnosis (Tsochatzidis et al 2021, Lee et al 2020) and treatment planning. With regard to tumors, such as lung cancer (Xue et al 2021), lymphoma and melanoma (Hatt et al 2017), are easily identified as the region with high fluorodeoxyglucose (FDG) uptake after the injection of 18F-FDG radiotracers. Although the PET tumor region has high standardized uptake values (SUVs), accurate PET/CT tumor segmentation still faces great challenges. For example, normal organs, such as the heart or sites of inflammation, can cause increased FDG uptake; conversely, in some well-differentiated lung tumors region, only a small amount of FDG is absorbed (Fu *et al* 2020). Therefore, the problem of over and under-segmentation will be introduced during tumor segmentation, which is also a challenge in this task.

Due to the high uptake of radiotracers at the tumor site, some methods use SUVs of the region of interest as a criterion for segmentation (Ilyas *et al* 2018, Eude *et al* 2021). These methods require to set a fixed threshold for the segmentation area, so knowledge of anatomy is required, and the resulting segmentation boundary is not particularly neat. In recent years, thanks to the powerful feature extraction ability of deep convolutional neural networks (CNN), it performs well in the segmentation task (Xing *et al* 2022, Zhang *et al* 2022). For example, UNet for biomedical segmentation has achieved satisfactory performance and become the most popular model. Based on the UNet framework, many different networks designed for different tasks are derived, such as VNet (Milletari *et al* 2016a), nnUNet (Isensee *et al* n.d.), UNet++ (Zhou *et al* 2020), and a series of variants (Wang *et al* 2022, 2021, Ni *et al* 2022). The VNet is modified to process 3D images on the basis of UNet, and introduces a jump connection in each stage, which better solves the problem of unbalanced foreground and background voxels. UNet++ redesigns the jump link so that the encoder can aggregate features of different scales, resulting in a highly flexible feature fusion scheme. At the same time, UNet++ can conduct deep supervision in different stages. Finally, nnUNet designs anisotropic kernel size and strides according to the data's coordinates. In addition, a residual connection was made in each convolution block.

In multi-modal PET/CT images, neural networks based on UNet structures also performed well in tumor segmentation (Blanc-Durand *et al* 2021, Jemaa *et al* 2020, Shiyam Sundar *et al* 2022) when jointly using PET and CT images instead of single PET image. Bi et al. (Bi *et al* 2021) proposed the reuse of multi-modality image features to refine tumor segmentation results. Jemaa et al. (Jemaa *et al* 2020) proposed an end-to-end method combining 2D and 3D CNN for tumor segmentation of PET/CT. Blanc-Durand et al. (Blanc-Durand *et al* 2021) proposed to use a 3D CNN for lymphoma lesion detection and segmentation of whole-body PET/CT images.

Although the CNN method has improved the performance of the traditional method in segmentation tasks, long-distance information is not well modeled due to limited the receptive field of CNN. In whole-body tumor segmentation tasks, performing long-range interactions is an important factor for reasoning in accurate tumor segmentation. For example, CNN methods prone to assign voxels with SUVs similar in a local area to the same category. Meanwhile, voxels of a tumor distributed at a distance are difficult to construct dependencies.

Recently, graph-based methods (Zhang *et al* 2020, Pan *et al* 2021, Demir *et al* 2021, Soberanis-Mukul *et al* 2020, Zhai *et al* 2019, Tian *et al* 2021) have also achieved satisfactory performance in medical image tasks, which benefits in its global long range information connected. Jin et al. (Jin *et al* 2017) refined the discontinuous results generated by fully convolutional network by GCN, and made considerable improvements in intrathoracic airway segmentation. Garcia-Uceda Juarez et al. (Garcia-Uceda Juarez *et al* 2019) proposed replacing the deepest level of the UNet with a series of graph convolutions to provide information on the connectivity of deep features. This end-to-end deep learning approach achieves good results in the airway segmentation task from chest CT scans. Considering the Euclidean and non-Euclidean features, Zhai et al. (Zhai *et al* 2019) proposed an end-to-end network combining CNN and GCN for the classification of pulmonary vascular arteriovenous vessels. Tian et al. (Tian *et al* 2020) used GCN to predict the contour of multi-scale prostate segmentation features obtained by CNN and achieved more accurate segmentation results. Lu et al. (Lu *et al* 2020) used GCN method to solve the task of image semantic segmentation for the first time. The receptive field can be expanded while avoiding the loss of local location information which usually ignored by CNN methods. In order to use the global structure information and self-similar details, Hu et al. (Hu *et al* 2021) introduced the graph network structure, which took the sub-regions with similar features in the image as nodes and formed connections according to similarity. Li et al. (Li *et al* 2021) used GCN network to segment retinal optical coherence tomography (OCT) images. Wu et al. (Wu *et al* 2022) modeled the graph network with super-voxels as units in brain MRI images, so as to perform tissue segmentation, which has better segmentation performance compared with CNN results.

However, when establishing graph structure by a randomly initialized adjacency matrix, it lacked analysis of input data, and could not establish graph structure well. Different from previous work, we rely on the input data and CNN results to analyze and establish the adjacency matrix. Specifically, we propose a refined CNN results method based on GCN network, which makes use of the uncertainty of CNN segmentation model and GCN network to reduce the false positive rate. The main contributions of this work are as follows:

1) Refine CNN segmentation uncertain region through GCN. Firstly, the uncertainty analysis of the results of the nnUNet segmentation is carried out. The analyzed voxels are then divided into a training set and a test set. The training set label is the preliminary segmentation result, and the test set is the voxels to be refined.

2) Define a new edge selection method. After the training set is determined, edges are selected for GCN to build the graph. Not only the six neighbors of each voxel are selected as edges to obtain local information, but also the distant voxels in different connected domains are selected as edges to obtain global information.

3) Finally, the experimental results show that in the dataset of 30 randomly selected patients, our proposed method reduces the false positive rate and improves the accuracy of tumor segmentation.

The remainder of this paper is organized as follows. Section 2 presents the establishment of a tumor segmentation graph model, including node selection, edge selection, and graph network model. Section 3 describes of the experiment settings and the results of the experiments. Section 4 discusses some related issues of our proposed nnUNet-GCN method and its limitations. Section 5 concludes the paper.

## 2. Methods

### 2.1 Method overview

Inspired by Soberanis-Mukul's GCN refinement work (Soberanis-Mukul *et al* 2020), we propose a graph network post-processing method based on uncertainty analysis for tumor segmentation of whole-body PET/CT images. In (Soberanis-Mukul *et al* 2020), the set of nodes in the graph is determined by analyzing the results of a 2D UNet segmentation. The edges in the graph are composed of 6-neighborhoods of these nodes and 16 random other pixels. The pipeline architecture of the proposed method is shown in Fig. 1, in which the first part is the 3D PET/CT input data, the second part is the trained nnUNet network, and then the preliminary segmentation results of the third part are obtained. The nnUNet segmentation results are analyzed to obtain the nodes of the fourth part. The fourth part of the figure shows the enlargement of the red box in the preliminary results. The white part of the node selection section is the nodes where the CNN result is used as a label, and the orange point is the uncertainty point to be optimized. Section B will describe how to get the nodes that make up the graph. After selecting the nodes that make up the graph, the network structure of the edge generation graph is formed in the last part of Fig. 1. The green dots are labeled nodes and the orange dots are the nodes to be predicted. The blue line is a long-distance connection, and the orange line is a local connection. Section C will introduce the method of edge selection. It is worth noting that when constructing edges, adding edges between the uncertain part and other points enhanced the graph network's ability to predict uncertain nodes.

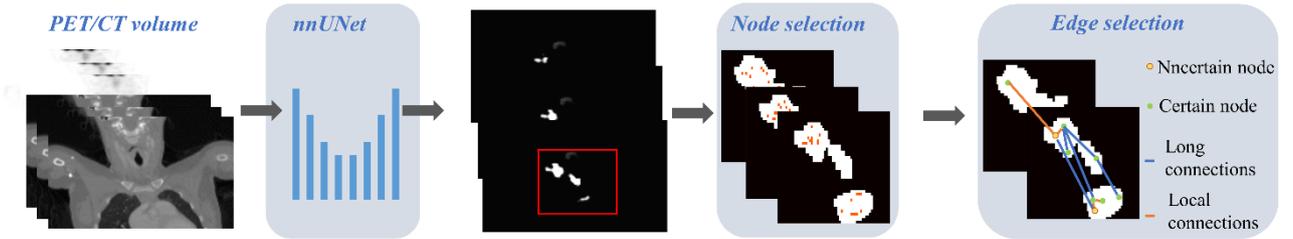

Fig.1. Overall pipeline of graph generaton. It consists of three components: a pre-trained nnUNet model, node selection, and edge selection par

### 2.2 Uncertainty analysis (Nodes selection)

We form the graph structure by selecting nodes and defining node connections. Let $\mathcal{G} = (\mathcal{V}, \mathcal{E})$ be a graph, where $\mathcal{V} = \{v_1, \dots, v_N\}$ is a collection of train and test nodes. $\mathcal{E} = \{A_{i,j} | v_i, v_j \in \mathcal{V}\}$ indicates a connection between $v_i$ and $v_j$ with a weight of $A_{i,j}$, called an edge. This section introduces the definition of $\mathcal{V}$, and the definition of $E$ is introduced in the next section

First, the input data $x$ obtains the foreground probability map through the trained nnUNet, as shown in Fig. 2 (c), which can be expressed as:

$$E(x) \approx f(x, \theta) \quad (1)$$

where $f$ is nnUNet is used in this paper. Uncertainty analysis was then performed on the predicted probability map of nnUNet. The entropy value is calculated by the foreground probability map serve as uncertainty evaluation criterion, as shown in Fig. 2 (d), which can be expressed as:

$$U(x) = -\sum_{c=1}^{M} E(x)^c \log E(x)^c \quad (2)$$

where $E(x)^c$ is the probability that the voxel $x$ belongs to category $C$ and $M$ is the category number to be divided, in which the value is 2 in this tumor segmentation task to distinguish normal regions from tumor. The uncertainty of a voxel is analyzed according to its entropy value, and its higher value indicates its higher uncertainty. Then set the threshold value of $U(x)$ to obtain the binary entropy value, as shown in Fig. 2 (f), which can be expressed as:

$$U_b(x) = U(x) > \alpha \quad (3)$$

in which voxels greater than $\alpha$ are used as uncertainty nodes for subsequent GCN refinement, shown as orange dots in the node selection part of Fig. 1. And $\alpha$ is set to 0.8 in this paper. After the test node is determined, the set threshold binarization operation is also performed on the probability map $E(x)$, as shown in Fig. 2 (e), which can be expressed as:

$$E_b(x) = E(x) > \beta \quad (4)$$

in which voxels bigger than $\beta$ are used as positive samples, and the label is 1 as GCN training data. In this article, the $\beta$ is set to 0.5. After the positive sample is determined, the dilation of $E_b(x) \cup U_b(x)$ yields the negative sample of the graph network with a label of 0. The training data for positive and negative samples contained in the graph network is shown in white part in Fig. 2 (g). As can be seen at the red arrow in Fig. 2, the difference between Fig 2 (c) and (h) pixels is represented in Fig. 2 (f), indicating that the nodes to be optimized is selected correctly.

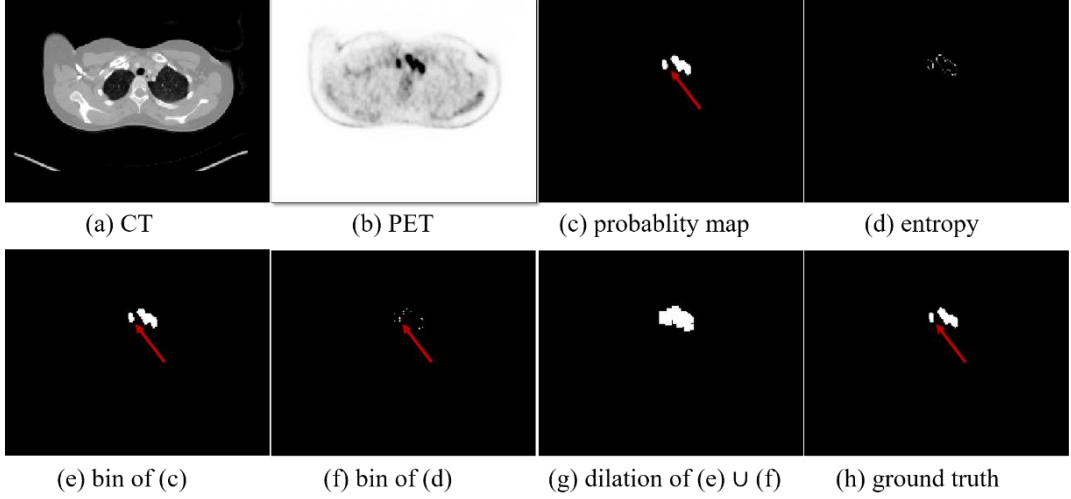

(a) CT    (b) PET    (c) probablity map    (d) entropy

(e) bin of (c)    (f) bin of (d)    (g) dilation of (e) ∪ (f)    (h) ground truth

Fig.2. Visual results of nodes selection process. (a) CT image, (b) PET image, (c) the probability map reslut of the nnUNet, (d) the entropy of (c), (e) binarized image of (c), (f) binarized image of (d), (g) dilation result of union of (e) and (f), and (h) the ground truth.

*2.3 Edges selection*

In this section, the definition of $\mathcal{E}$ will be introduced. In (Soberanis-Mukul *et al* 2020), two methods of establishing edges were used. In order to obtain local information, each node is connected to its nodes in 6 neighborhoods to form edges. The second method is to obtain global information, randomly selecting 16 nodes for each node to connect to form edges.

In this paper, since tumors exist in more than one site, to strengthen the information between distant tumors, we divide the preliminary segmentation probability results of nnUNet tumors into three parts. We take three ways to establish edges to obtain information between tumors. First, the voxels of each part of the tumor establish edges with its 6-neighbors to obtain local tumor information. Second, edges are established between the voxels within each part of the tumor and the voxels in other parts of the tumor to obtain global tumor information. Third, since the voxels with high uncertainty are the nodes to be optimized, it is necessary to select nodes to establish edges in different parts of the tumor region to obtain more information.

In (Soberanis-Mukul *et al* 2020), spatial information as well as intensity are used as weight information for edges. Because there is a gap between the uncertainty nodes and the certain nodes (tumor and non-tumor regions) in intensity, the results are weighted less and bring less information. Furthermore, because the distance information is used as a weight coefficient, it will focus on local information and weaken long-distance information. For these reasons, we assign each edge a weight of 1. After obtaining graph $\mathcal{G}$, the GCN network was trained by Kipf's method (Kipf and Welling 2017), and then the uncertainty nodes were predicted to obtain the tumor segmentation results after refinement.

*2.4 GCN refinement*

After selecting nodes from Section 2.2 and establishing edges in Section 2.3, a graph convolutional layer can be defined as:

$$H = \sigma(\widetilde{D}^{-\frac{1}{2}}\widetilde{A}\widetilde{D}^{-\frac{1}{2}}XW), \quad \widetilde{A} = A + I_N \quad (5)$$

where $\sigma(\cdot)$ is rectified Linear Unit, $A$ is the adjacency of graph $\mathcal{G}$, $I_N$ is a unit matrix and $\widetilde{D}$ is the diagonal matrix. $X \in \mathcal{R}^{N \times K}$ is the input data, $N$ is the nodes of $\mathcal{G}$, in this case, the voxels with high confidence labeled with the nnUNet result (containing the tumor region, labeled 1, and the non-tumor region, labeled 0), and high uncertainty unlabeled voxels. $K$ is the feature of each node, and in this paper, each node has 4 features, which are PET/CT voxel value, nnUNet prediction probability $P(x)$, and entropy value $U(x)$. $W \in \mathcal{R}^{K \times K'}$ is the weight matrix to be optimized, and receiving $K$ features yields the output $K'$ features. Finally, referring to the model structure of (23), a two-layer simple GCN model is obtained:

$$Y = \Gamma(A, X; W^0, W^1) = sigmoid\big(H_2(A, H_1(A, X; W^0); W^1)\big) \quad (6)$$

**3. Experiments**

*3.1 Setup*

We validate our GCN refining method on nnUNet 3D tumor segmentation results. The refined method in this paper is compared with the nnUNet method. Then, we performed an ablation study of different ways to build edges to compare the impact on the results. The nnUNet takes Dice as the loss function, trains with Adam optimizer, takes 200 PET/CT data with tumors as the training set and 20 data as the validation set, and stopped training when the validation set loss function no longer decreases. Finally, 30 cases were randomly selected as the testing set to verify the experiments in this paper.

*3.1.1 Data.* In the experiment, we used PET/CT data from the MICCIA2022 autoPET challenge (Gatidis *et al* 2022), a dataset containing 1014 sets of training data from 900 patients at University Hospital Tubingen, Germany. The scans included malignant melanoma, lymphoma, or lung cancer as well as negative control patients. We randomly selected 220 scans with tumors from the dataset as nnUNet training and validation data, and then randomly selected 30 scans with tumors as test sets to evaluate and verify the proposed GCN refinement method.

*3.1.2 Evaluation metric.* To quantitatively evaluate the segmentation performance of different methods, the Dice similarity coefficient (Dice), 95% Hausdorff distance (HD95), and average symmetric surface distance (ASSD) were used to quantitatively evaluate tumor segmentation performance. Dice evaluates the overlap ratio between the segmentation result and the true label at the pixel level, and a higher value indicates that the segmentation result is closer to the true label. HD95 and ASSD are used to calculate the average distance of the surface that splits two 3D volumes, and a lower value indicates that the boundaries of the 3D segmentation region are closer, having a better segmentation performance.

*3.2 Comparison to state-of-the-art methods*

In this section, we compared the proposed method, UNet, UNet+GCN, VNet, SegResNet and nnUNet methods quantitatively and qualitatively. In the quantitative analysis, the experimental results can be seen in Table 1. In terms of Dice score, UNet, VNet, UNet+GCN (baseline), SegResNet, and nnUNet 3D method obtained an average of 65.48%, 66.89%, 66.90%, 68.82%, 72.73% and 74.51% results in the test set, respectively. Our method achieved an average score of 76.63%, which improved by 2.12 percentage based on nnUNet segmentation results, indicating that our post-treatment method is closer to the real tumor in terms of segmentation results. Notably, the proposed method has a 7.81% Dice improvement compared with the baseline UNet+GCN method, indicating that after uncertainty analysis of nnUNet results, adding edges between uncertain nodes and other nodes can make the graph network more accurately distinguish whether uncertain nodes belong to tumor regions. In HD95 and ASSD, the best performance of 31.63 and 4.77, outperformed the nnUNet method with 6.34 and 1.72, respectively. It shows that the segmentation results after post-processing of our graph network are significantly improved compared with the state-of-the-art methods.

Table 1: Quantitative results of 30 test data with different methods. The average Dice (%), HD95 and ASSD value are shown.

| Method | Dice(%)↑ | HD95↓ | ASSD↓ |
|---|---|---|---|
| UNet (Ronneberger *et al* 2015) | 65.48 | 69.61 | 12.41 |
| VNet (Milletari *et al* 2016b) | 66.89 | 50.39 | 10.64 |
| UNet-GCN Refinement (Soberanis-Mukul *et al* 2020) | 68.82 | 37.85 | 7.67 |
| SegResNet (Myronenko 2018) | 72.73 | 40.13 | 6.64 |
| nnUNet (Isensee *et al* 2018) | 74.51 | 37.97 | 6.49 |
| nnUNet-GCN Refinement (proposed) | **76.63** | **31.63** | **4.77** |

In order to comprehensively demonstrate the performance of tumor segmentation using different methods in the test set, three tumors of different sizes were selected for visualization, and the results are shown in Fig 3, 4 and 5. Fig. 3 shows the visual segmentation results with different segmentation methods for patients with small tumors. Red, green and blue represent false positive, true positive and false negative, respectively. As can be seen from the first row of Fig. 3, CNN and CNN+GCN methods have false positive and false negative results. The Vnet, SegResNet and nnUNet methods improve the false positive situation, but there are still false positive results, and the false positive results can be effectively removed by our proposed nnUNet+GCN method. Also in the second row, it can be seen that the proposed nnUNet+GCN method removes false positive and has better performance.

The intuitive 3D perspective segmentation results are shown in the third row of Fig. 3, and it can be seen that the green part of our proposed method occupies more of it, indicating closer to ground truth. It can be seen that our nnUNet+GCN performs well when dealing with long-distance small over-segmentation of tumors, indicating that we combine long-distance information when establishing edges.

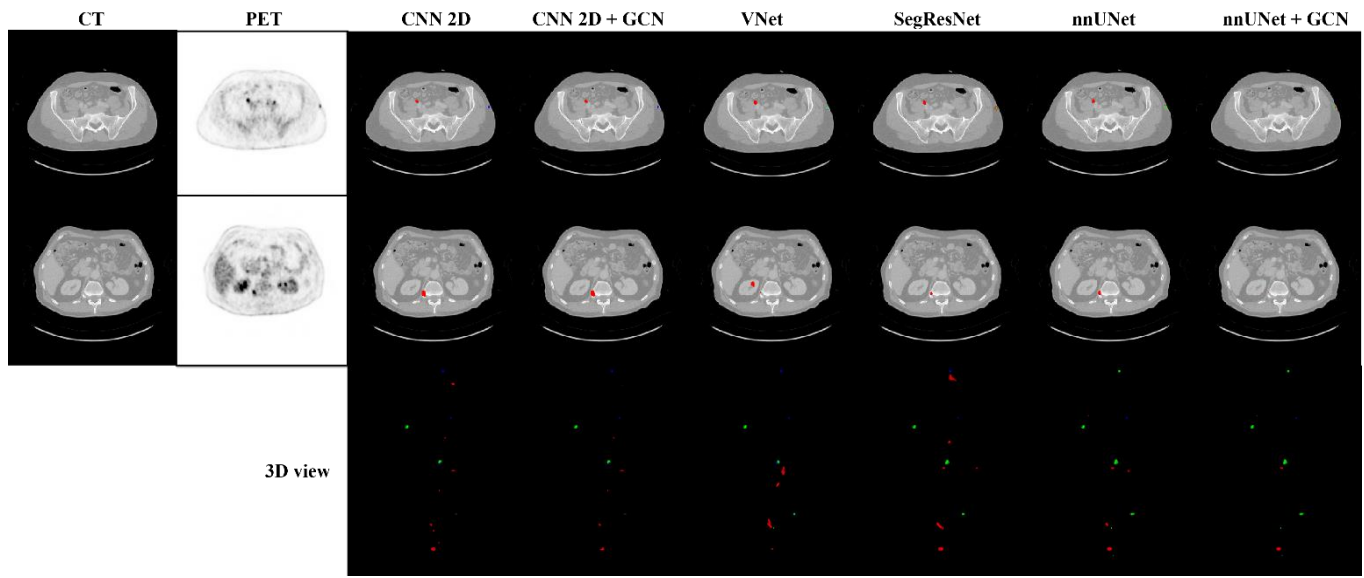

Fig. 3. Comparison of different method segmentation results with small tumors. The first two columns are CT and PET images, the following six columns show the segmentation results with CNN 2D, CNN 2D + GCN, VNet, SegResNet, nnUNet and nnUNet + GCN methods. The first two raws are two slices in the axial direction, and the third row is the 3D view results. Blue, green and red delineations means false positive, true positive and false negative, respectively.

Fig. 4 shows the segmentation results of another case, as can be seen from the first column of the figure, CNN 2D, CNN 2D+GCN and VNet methods do not recognize high FDG uptake locations as tumors on the pulmonary arteries, while SegResNet and nnUNet mistakenly identify them as tumors, and the proposed nnUNet+GCN method can remove a part of the false positives. As can be seen from the 3D view in the second row, while CNN 2D, CNN 2D+GCN, and VNet perform well on the false positive problem, the blue part indicates that there is an obvious false negative problem, while SegResNet and nnUNet have a false positive problem.。 In the results of nnUNet segmentation, it can be clearly seen that the location at a distance below is incorrectly identified as a tumor, but the proposed nnUNet+GCN method suppresses nnUNet false positive result, and obtains a more realistic segmentation result.

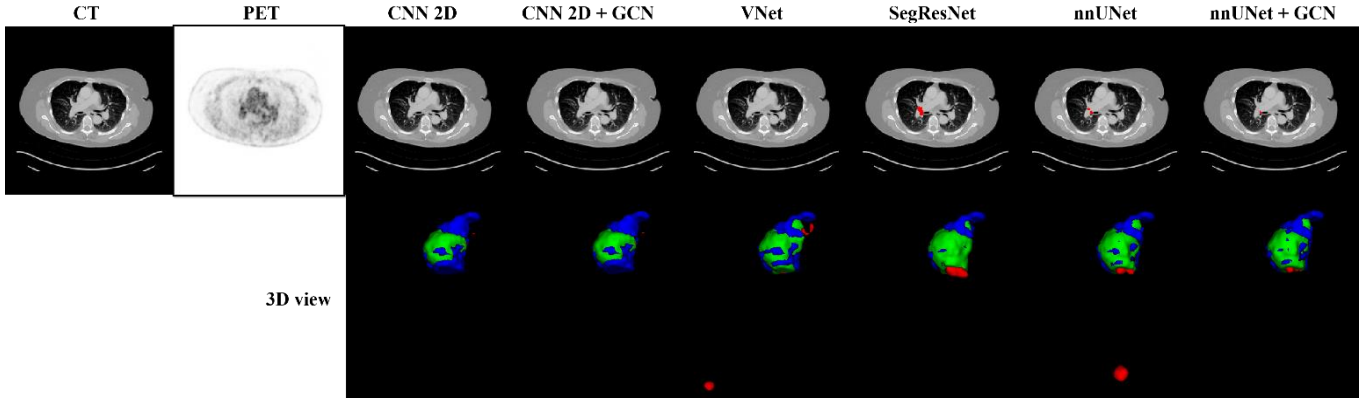

Fig. 4. Comparison of different method segmentation results with medium-size tumors. The first two columns are CT and PET images, the following six columns show the segmentation results with CNN 2D, CNN 2D + GCN, VNet, SegResNet, nnUNet and nnUNet + GCN methods. The first raw is slice in the axial direction, and the third row is the 3D view results. Blue, green and red delineations means false positive, true positive and false negative, respectively.

Fig. 5 shows the results of the large tumor case segmentation experiment. As can be seen from the third row to column 8 of the first column in the figure, all methods have a false negative problem. The CNN 2D segmentation results have a false positive problem, but it can be seen from the fourth column that the red part disappears after GCN refine, indicating that the false positive problem of the CNN 2D segmentation method has been improved. The second row of Fig. 5 shows the 3D perspective of different segmentation results, and it can be seen that the CNN2D and CNN2D+GCN methods have more blue parts, indicating that there are too many false negative problems. A large portion of the results of VNet, SegResNet, and nnUNet splits are red, indicating a false positive problem with their methods. We can see that even if there is a very large over-segmentation situation, it can be removed after the proposed nnUNet + GCN post-processing method is optimized.

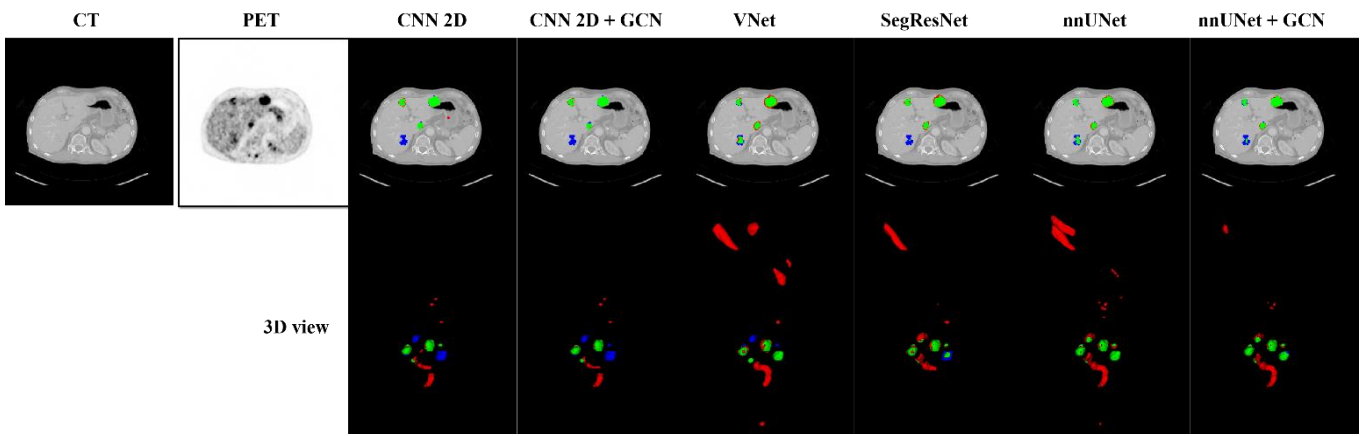

Fig. 5. Comparison of different method segmentation results with large tumors. The first two columns are CT and PET images, the following six columns show the segmentation results with CNN 2D, CNN 2D+GCN, VNet, SegResNet, nnUNet and nnUNet + GCN methods. The first raw is slice in the axial direction, and the third row is the 3D view results. Blue, green and red delineations means false positive, true positive and false negative, respectively.

*3.3 Ablation study*

In order to investigate the effects of edge selection, we perform the ablations on this process in different ways of establishing edges. We show the quantitative results obtained by different ways of constructing edges in the 3-5 rows of Table 2. The first to third columns are three different ways to select edges. 6n+16r means that each node establishes an edge with 22 nodes, which contain 6 neighborhood nodes and 16 random other nodes. To improve the accuracy of the prediction of the test node (orange dot in the node selection part in Fig. 1.), we add nodes formation edges to the test nodes. There are two methods, one is to pick the node build edge in the training set, which we write as 16uncer-cer. The other is to randomly select the node construction edge, which is denoted as 16uncer-rand. We can see the result of the baseline method for the third row. The fourth is to add 32 edges between test and train nodes. The fifth row is the result of adding 32 edges between the test node and random other nodes. It can be seen from the results, adding edges to the test nodes, can be significantly improved in quantitative indicators. Two different methods of increasing edges improved by 0.21% and 0.27% in average Dice scores, respectively. The lowest values were achieved on both the HD95 and ASSD indicators, which further proved the superiority of the proposed method.

Table 2. Quantitative results of different edge selection methods for GCN based on nnUNet segmentation. The average Dice (%), HD95 and ASSD value are shown.

| Method | | | Metric | | |
| --- | --- | --- | --- | --- | --- |
| 6n+16r | 16uncer-cer | 16uncer-rand | Dice(%)↑ | HD95↓ | ASSD↓ |
| √ | – | – | 76.28 | 32.64 | 4.93 |
| √ | √ | – | 76.58 | 31.66 | 4.80 |
| √ | – | √ | **76.63** | **31.63** | **4.77** |

## 4. Discussions

In this work, we propose a PET/CT tumor segmentation post-segmentation refinement method, which analyzes the initial segmentation results of nnUNet and forms a graph to reclassify the uncertain nodes. nnUNet has the problem of over-segmentation when segmenting PET/CT images. For example, in the 3D view of Fig. 3, 4 and 5, the regions of red parts are easily identified as tumors. In order to solve this problem, a method for optimizing nnUNet results based on GCN is proposed. The proposed method analyzes the initial segmentation results according to the entropy value of the probability map to obtain the certain nodes and uncertain nodes, and the uncertain nodes are the nodes to be optimized. When building the graph, we divide the nodes into three parts, and in each part nodes, we add uncertain nodes to form edges with nodes from other parts to get more mutual information.

In quantitative results, the average Dice coefficient is increased by 2.12% compared to the nnUNet results, which can effectively suppress false-positive tumors in small, medium and large regions. In addition, the proposed method yielded lower value in HD95 as well as the ASSD indicator. We also show different ways to establish edges. It can be seen from the experimental results that the GCN prediction ability can be effectively improved by increasing the edges for the uncertainty nodes.

This work still has some limitations. First, although the proposed method can effectively reduce the false positive rate and improve the segmentation accuracy, there is a potential possibility that the erroneous result in nnUNet will be regarded as the correct result in the uncertainty analysis, which will bring incorrect sample labels to subsequent GCN training, so there is still a problem of over-segmentation. Second, the proposed method is a post-processing method, which does not process the input data and lacks analysis of the tumor under-segmentation part of the nnUNet network. In future work, we consider combining GCN network with CNN network for tumor segmentation of 3D PET/CT images. The nnUNet over-segmentation problem is solved by supervised learning training and the input data is directly processed to solve the under-segmentation problem of nnUNet network.

## 5. Conclusions

In this work, we propose a sparse semi-supervised graph network for reducing the false positive rate of 3D PET/CT tumor segmentation, which is constructed on the 3D nnUNet tumor segmentation results. The nnUNet + GCN were evaluated in 30 randomly selected data, showing that the segmentation accuracy was improved with GCN refinement. Both quantitative and qualitative results indicate that our proposed PET/CT tumor segmentation method is superior to UNet, UNet-GCN, VNet, SegResNet and nnUNet methods.

**Acknowledgements**

This work was supported by the Fundamental Research Funds for the Central Universities, China (N2019006 and N180719020).

**References**

Bi L, Fulham M, Li N, Liu Q, Song S, Dagan Feng D and Kim J 2021 Recurrent feature fusion learning for multi-modality pet-ct tumor segmentation *Computer Methods and Programs in Biomedicine* **203** 106043

Blanc-Durand P, Jégou S, Kanoun S, Berriolo-Riedinger A, Bodet-Milin C, Kraeber-Bodéré F, Carlier T, Le Gouill S, Casasnovas R-O, Meignan M and Itti E 2021 Fully automatic segmentation of diffuse large B cell lymphoma lesions on 3D FDG-PET/CT for total metabolic tumour volume prediction using a convolutional neural network *Eur J Nucl Med Mol Imaging* **48** 1362–70

Demir U, Ozer A, Sahin Y H and Unal G 2021 Uncertainty-Based Dynamic Graph Neighborhoods For Medical Segmentation Online: http://arxiv.org/abs/2108.03117

Eude F, Toledano M N, Vera P, Tilly H, Mihailescu S-D and Becker S 2021 Reproducibility of Baseline Tumour Metabolic Volume Measurements in Diffuse Large B-Cell Lymphoma: Is There a Superior Method? *Metabolites* **11** 72

Fu X, Bi L, Kumar A, Fulham M and Kim J 2020 Multimodal Spatial Attention Module for Targeting Multimodal PET-CT Lung Tumor Segmentation Online: http://arxiv.org/abs/2007.14728v1

Garcia-Uceda Juarez A, Selvan R, Saghir Z and de Bruijne M 2019 A Joint 3D UNet-Graph Neural Network-Based Method for Airway Segmentation from Chest CTs *Machine Learning in Medical Imaging* Lecture Notes in Computer Science vol 11861, ed H-I Suk, M Liu, P Yan and C Lian (Cham: Springer International Publishing) pp 583–91 Online: http://link.springer.com/10.1007/978-3-030-32692-0_67

Gatidis S, Hepp T, Früh M, La Fougère C, Nikolaou K, Pfannenberg C, Schölkopf B, Küstner T, Cyran C and Rubin D 2022 A whole-body FDG-PET/CT Dataset with manually annotated Tumor Lesions *Sci Data* **9** 601

Hatt M, Tixier F, Pierce L, Kinahan P E, Le Rest C C and Visvikis D 2017 Characterization of PET/CT images using texture analysis: the past, the present… any future? *Eur J Nucl Med Mol Imaging* **44** 151–65

Hu X, Yan Y, Ren W, Li H, Bayat A, Zhao Y and Menze B 2021 Feedback Graph Attention Convolutional Network for MR Images Enhancement by Exploring Self-Similarity Features *Proceedings of the Fourth Conference on Medical Imaging with Deep Learning* Medical Imaging with Deep Learning (PMLR) pp 327–37 Online: https://proceedings.mlr.press/v143/hu21a.html

Ilyas H, Mikhaeel N G, Dunn J T, Rahman F, Møller H, Smith D and Barrington S F 2018 Defining the optimal method for measuring baseline metabolic tumour volume in diffuse large B cell lymphoma *Eur J Nucl Med Mol Imaging* **45** 1142–54

Isensee F, Petersen J, Klein A, Zimmerer D, Jaeger P F, Kohl S, Wasserthal J, Koehler G, Norajitra T, Wirkert S and Maier-Hein K H 2018 nnU-Net: Self-adapting Framework for U-Net-Based Medical Image Segmentation Online: http://arxiv.org/abs/1809.10486

Isensee F, Petersen J, Klein A, Zimmerer D, Jaeger P F, Kohl S, Wasserthal J, Kohler G, Norajitra T and Maier-Hein K H nnU-Net: Self-adapting Framework for U-Net-Based Medical Image Segmentation 12

Jemaa S, Fredrickson J, Carano R A D, Nielsen T, de Crespigny A and Bengtsson T 2020 Tumor Segmentation and Feature Extraction from Whole-Body FDG-PET/CT Using Cascaded 2D and 3D Convolutional Neural Networks *J Digit Imaging* **33** 888–94

Jin D, Xu Z, Harrison A P, George K and Mollura D J 2017 3D Convolutional Neural Networks with Graph Refinement for Airway Segmentation Using Incomplete Data Labels *Machine Learning in Medical Imaging* Lecture Notes in Computer Science vol 10541, ed Q Wang, Y Shi, H-I Suk and K Suzuki (Cham: Springer International Publishing) pp 141–9 Online: http://link.springer.com/10.1007/978-3-319-67389-9_17

Kipf T N and Welling M 2017 Semi-Supervised Classification with Graph Convolutional Networks Online: http://arxiv.org/abs/1609.02907

Lee J W, Park Y-J, Jeon Y S, Kim K H, Lee J E, Hong S H, Lee S M and Jang S J 2020 Clinical value of dual-phase F-18 sodium fluoride PET/CT for diagnosing bone metastasis in cancer patients with solitary bone lesion *Quantitative Imaging in Medicine and Surgery* **10** 2098111–2092111

Li J, Jin P, Zhu J, Zou H, Xu X, Tang M, Zhou M, Gan Y, He J, Ling Y and Su Y 2021 Multi-scale GCN-assisted two-stage network for joint segmentation of retinal layers and disc in peripapillary OCT images Online: http://arxiv.org/abs/2102.04799

Lu Y, Chen Y, Zhao D and Chen J 2020 Graph-FCN for image semantic segmentation Online: http://arxiv.org/abs/2001.00335

Milletari F, Navab N and Ahmadi S A 2016a V-Net: Fully Convolutional Neural Networks for Volumetric Medical Image Segmentation *2016 Fourth International Conference on 3D Vision (3DV)* Online: http://ieeexplore.ieee.org/document/7785132/

Milletari F, Navab N and Ahmadi S-A 2016b V-Net: Fully Convolutional Neural Networks for Volumetric Medical Image Segmentation *2016 Fourth International Conference on 3D Vision (3DV)* 2016 Fourth International Conference on 3D Vision (3DV) (Stanford,

CA, USA: IEEE) pp 565–71 Online: http://ieeexplore.ieee.org/document/7785132/

Myronenko A 2018 3D MRI brain tumor segmentation using autoencoder regularization Online: http://arxiv.org/abs/1810.11654

Ni Y, Xie Z, Zheng D, Yang Y and Wang W 2022 Two-stage multitask U-Net construction for pulmonary nodule segmentation and malignancy risk prediction *Quantitative Imaging in Medicine and Surgery* **12** 29209–309

Pan S-Y, Lu C-Y, Lee S-P and Peng W-H 2021 Weakly-Supervised Image Semantic Segmentation Using Graph Convolutional Networks Online: http://arxiv.org/abs/2103.16762

Ronneberger O, Fischer P and Brox T 2015 U-Net: Convolutional Networks for Biomedical Image Segmentation Online: http://arxiv.org/abs/1505.04597

Shiyam Sundar L K, Yu J, Muzik O, Kulterer O, Fueger B J, Kifjak D, Nakuz T, Shin H M, Sima A K, Kitzmantl D, Badawi R D, Nardo L, Cherry S R, Spencer B A, Hacker M and Beyer T 2022 Fully-automated, semantic segmentation of whole-body 18F-FDG PET/CT images based on data-centric artificial intelligence *J Nucl Med* jnumed.122.264063

Soberanis-Mukul R D, Navab N and Albarqouni S 2020 An Uncertainty-Driven GCN Refinement Strategy for Organ Segmentation Online: http://arxiv.org/abs/2012.03352

Tian Z, Li X, Chen Z, Zheng Y, Fan H, Li Z, Li C and Du S 2021 Interactive prostate MR image segmentation based on ConvLSTMs and GGNN *Neurocomputing* **438** 84–93

Tian Z, Li X, Zheng Y, Chen Z, Shi Z, Liu L and Fei B 2020 Graph-convolutional-network-based interactive prostate segmentation in MR images *Med Phys* **47** 4164–76

Tsochatzidis L, Koutla P, Costaridou L and Pratikakis I 2021 Integrating segmentation information into CNN for breast cancer diagnosis of mammographic masses *Computer Methods and Programs in Biomedicine* **200** 105913

Wang X, Bao N, Xin X, Tan J, Li H, Zhou S and Liu H 2022 Automatic evaluation of endometrial receptivity in three-dimensional transvaginal ultrasound images based on 3D U-Net segmentation *Quantitative Imaging in Medicine and Surgery* **12** 4095108–4094108

Wang Y, Zhang Y, Wen Z, Tian B, Kao E, Liu X, Xuan W, Ordovas K, Saloner D and Liu J 2021 Deep learning based fully automatic segmentation of the left ventricular endocardium and epicardium from cardiac cine MRI *Quantitative Imaging in Medicine and Surgery* **11** 1600612–1612

Wu C, Feng Z, Zhang H and Yan H 2022 Graph Neural Network and Superpixel Based Brain Tissue Segmentation *2022 International Joint Conference on Neural Networks (IJCNN)* 2022 International Joint Conference on Neural Networks (IJCNN) (Padua, Italy: IEEE) pp 01–8 Online: https://ieeexplore.ieee.org/document/9892580/

Xing H, Zhang X, Nie Y, Wang S, Wang T, Jing H and Li F 2022 A deep learning-based post-processing method for automated pulmonary lobe and airway trees segmentation using chest CT images in PET/CT *Quantitative Imaging in Medicine and Surgery* **12** 4747757–4744757

Xue H, Zhang Q, Zou S, Zhang W, Zhou C, Tie C, Wan Q, Teng Y, Li Y, Liang D, Liu X, Yang Y, Zheng H, Zhu X and Hu Z 2021 LCPR-Net: low-count PET image reconstruction using the domain transform and cycle-consistent generative adversarial networks *Quantitative Imaging in Medicine and Surgery* **11** 74962–74762

Zhai Z, Staring M, Zhou X, Xie Q, Xiao X, Els Bakker M, Kroft L J, Lelieveldt B P F, Boon G J A M, Klok F A and Stoel B C 2019 Linking Convolutional Neural Networks with Graph Convolutional Networks: Application in Pulmonary Artery-Vein Separation *Graph Learning in Medical Imaging* Lecture Notes in Computer Science vol 11849, ed D Zhang, L Zhou, B Jie and M Liu (Cham: Springer International Publishing) pp 36–43 Online: http://link.springer.com/10.1007/978-3-030-35817-4_5

Zhang L, Li X, Arnab A, Yang K, Tong Y and Torr P H S 2020 Dual Graph Convolutional Network for Semantic Segmentation Online: http://arxiv.org/abs/1909.06121

Zhang Y, Zhang D, Chen Z, Wang H, Miao W and Zhu W 2022 Clinical evaluation of a novel atlas-based PET/CT brain image segmentation and quantification method for epilepsy *Quantitative Imaging in Medicine and Surgery* **12** 4538548–4534548

Zhou Z, Siddiquee M M R, Tajbakhsh N and Liang J 2020 UNet++: Redesigning Skip Connections to Exploit Multiscale Features in Image Segmentation Online: http://arxiv.org/abs/1912.05074